\documentclass[12pt]{article}
\usepackage{epsfig,amssymb,amsmath,psfrag}

\def \bl  {\begin{align*}}
\def \el  {\end{align*}}

\def \be  {\begin{equation}}
\def \ee  {\end{equation}}
\def \ba  {\begin{eqnarray}}
\def \ea  {\end{eqnarray}}
\def \baa {\begin{eqnarray*}}
\def \eaa {\end{eqnarray*}}
\def \bb  {\begin {thebibliography} }
\def \eb  {\end{thebibliography}}
\def \lab #1 {\label{#1}}

\def \qqquad {\qquad\quad}

\newcommand{\beq}{\begin{equation}}
\newcommand{\eeq}{\end{equation}}
\newcommand{\beqa}{\begin{eqnarray}}
\newcommand{\eeqa}{\end{eqnarray}}

\newcommand{\tlam}{\tilde{\lambda}}
\def \e  {\mathop{\rm e}\nolimits}

\renewcommand{\a}{\alpha}
\newcommand{\adt}{{\dot{\alpha}}}
\newcommand{\bdt}{{\dot{\beta}}}
\renewcommand{\b}{\beta}

\def\e{\epsilon}
\def\d{\delta}
\def\lam{\lambda}
\def\tlam{\tilde{\lambda}}
\def\th{\theta}
\def\cN{\mathcal{N}}
\def\cP{\mathcal{P}}
\def\cW{\mathcal{W}}
\def\cZ{\mathcal{Z}}
\def\AA{\mathcal{A}}
\def\BB{\mathcal{B}}
\def\CC{\mathcal{C}}
\def\DD{\mathcal{D}}
\def\EE{\mathcal{E}}
\def\FF{\mathcal{F}}
\def\GG{\mathcal{G}}
\def\HH{\mathcal{H}}
\def\MM{\mathcal{M}}
\def\OO{\mathcal{O}}
\def\UU{\mathcal{U}}
\def\NN{\mathcal{N}}
\def\tW{\tilde{W}}

\def\del{\partial}

\def\l<{\langle}
\def\r>{\rangle}

\def\XXint#1#2#3{{\setbox0=\hbox{$#1{#2#3}{\int}$}
     \vcenter{\hbox{$#2#3$}}\kern-.5\wd0}}

\renewcommand{\title}[1]{\vbox{\center\LARGE{#1}}\vspace{5mm}}
\renewcommand{\author}[1]{\vbox{\center#1}\vspace{5mm}}

\begin{document}

% Front page here
\thispagestyle{empty}
\null\vskip-12pt \hfill
LAPTH-013/10
\\
\vskip2.2truecm
\begin{center}
\vskip 0.2truecm {\Large\bf
%\titleline
{\Large The Yangian origin of the Grassmannian integral}
}\\
\vskip 1truecm
{\bf J.~M. Drummond and L. Ferro \\
}

\vskip 0.4truecm
%$^{*}$
{\it
LAPTH\footnote{Laboratoire d'Annecy-le-Vieux de Physique Th\'{e}orique, UMR 5108}, Universit\'{e} de Savoie, CNRS\\
B.P. 110,  F-74941 Annecy-le-Vieux Cedex, France\\
\vskip .2truecm                        }
\end{center}

\vskip 1truecm %\Large
%\noindent
\centerline{\bf Abstract} % \normalsize
In this paper we analyse formulas which reproduce leading singularities of scattering amplitudes in $\NN=4$ super Yang-Mills theory through a Grassmannian integral.
Recently their Yangian invariance has been proved directly by using the explicit expression of the Yangian level-one generators. The specific cyclic structure of the form integrated over the Grassmannian enters in a crucial way in demonstrating the symmetry. Here we show that the Yangian symmetry fixes this structure uniquely.

\medskip

 \noindent

\newpage
\setcounter{page}{1}\setcounter{footnote}{0}
%\tableofcontents
%\newpage

%%%%%%%%%%%%%%%%%%%%%%%%%%%%%%%%%%%%%%%%%%%%%%%%
\section{Introduction}

In recent years much progress has been made in the study of scattering amplitudes in planar $\NN = 4$ supersymmetric Yang-Mills theory. It is known that this theory possesses a very large symmetry in its planar limit. Indeed the theory seems to have some underlying integrable structure which governs all the physical quantities. This has been seen at the level of the spectrum of anomalous dimensions \cite{AdS/INT1,AdS/INT1b,Beisert:2005fw}. It is related to the fact that planar $\NN=4$ super Yang-Mills theory is also described by the IIB superstring on an AdS${}_5 \times$S${}^5$ background \cite{Maldacena:1997re,Gubser:1998bc,Witten:1998qj} and this theory is classically integrable \cite{Bena:2003wd}.

The integrable structure also shows up in the scattering amplitudes of the planar theory. Indeed it has been discovered recently that they are constrained by a hidden symmetry which is not present in the Lagrangian of the theory. Indeed planar scattering amplitudes in $\NN=4$ SYM are related to light-like Wilson loops  \cite{Alday:2007hr,Drummond:2007aua,Brandhuber:2007yx,Drummond:2007cf,Drummond:2007au,Drummond:2007bm,Drummond:2008aq}.
The Wilson loops live in a dual coordinate space, defined by $p_i = x_i - x_{i+1}$ \cite{Drummond:2006rz} where the $p_i$ are incoming momenta of the particles in the scattering process. The natural conformal symmetry of the Wilson loops therefore acts on the scattering amplitudes. This new dual conformal symmetry is distinct from the original conformal symmetry of the Lagrangian.
Its extension to dual superconformal symmetry is very natural and it partially overlaps with the original superconformal symmetry \cite{Drummond:2008vq,Berkovits:2008ic,Beisert:2008iq}.
The dual conformal symmetry is broken by loop corrections, but in a controlled way \cite{Drummond:2007cf,Drummond:2007au,Drummond:2008vq} (see \cite{Alday:2009zm} for a regularisation in which the symmetry is restored).
At tree-level, however the symmetry is unbroken \cite{Drummond:2008vq,Brandhuber:2008pf}. Indeed the explicit solution of the BCFW recursion relations \cite{Britto:2004ap,Britto:2005fq} for all tree-level amplitudes in \cite{Drummond:2008cr} is written in terms of dual superconformal invariants.

In \cite{Drummond:2009fd} the combination of the ordinary superconformal and dual superconformal symmetries has been shown to form a Yangian symmetry in the bilocal representation described in \cite{Dolan:2003uh,Dolan:2004ps}.  The level-zero subalgebra of the Yangian is provided by the original superconformal symmetry while the bilocal level-one generators arise from combining the original symmetry with the dual conformal generators. At tree-level there are no infrared divergences and so the original superconformal symmetry of the Lagrangian is unbroken except on singular kinematic configurations \cite{Bargheer:2009qu,Korchemsky:2009hm,Sever:2009aa}. This subtle non-invariance of the tree amplitudes implies a more severe symmetry breaking for loop corrections \cite{Korchemsky:2009hm,Sever:2009aa,Beisert:2010gn}. The breaking can be understood by deforming the tree-level generators so that they annihilate the one-loop amplitude \cite{Beisert:2010gn}.
Indeed the original superconformal symmetry of simple BCFW terms can be verified directly \cite{Korchemsky:2009jv}. This property is transparent when the BCFW recursion is formulated in twistor space \cite{ArkaniHamed:2009si,Mason:2009sa} so each of the terms in the BCFW expansion at tree-level is in fact a Yangian invariant.

The terms in the tree-level amplitude are also present at one-loop as box integral coefficients \cite{Britto:2004nc,Brandhuber:2008pf,Drummond:2008bq,ArkaniHamed:2008gz}. This is necessary \cite{Bern:2004bt} in order to be consistent with the known factorisation of infrared divergences in gauge theory. There are other box-integral coefficients at one-loop, the four-mass coefficients, which do not appear in the tree-level amplitude. Likewise there are higher-loop leading singularities (see \cite{Cachazo:2008vp} for a discussion of leading singularities) which do not appear at tree-level or one loop.

Recently, a remarkable formula was proposed in \cite{ArkaniHamed:2009dn} that seems to capture all of these objects in one go. The formula is an integral over the Grassmannian $G(k,n)$ of a specific cyclic $k(n-k)$-form built from superconformally invariant delta functions of linear combinations of twistor variables. It has been conjectured that this object captures all of the leading singularities of $\NN=4$ super Yang-Mills amplitudes, with explicit evidence being given in \cite{ArkaniHamed:2009dn} for amplitudes with a low number of external legs. The leading singularities arise by choosing different contours of integration in Grassmannian.
Even though each choice of contour appears to produce a dual superconformally invariant leading singularity, the dual superconformal symmetry is not immediately apparent in the initial integral formula.
Many developments based on this formula have been pursued \cite{Bullimore:2009cb,Kaplan:2009mh,ArkaniHamed:2009sx,ArkaniHamed:2009dg}. In particular a T-dual version was proposed \cite{Mason:2009qx} which is of exactly the same form but in terms of momentum twistors \cite{Hodges:2009hk}, the twistors associated to the dual coordinate space. This form thus has dual superconformal symmetry manifest while the original superconformal symmetry is not obvious. The two formulas were in fact shown to be equivalent to each other in \cite{ArkaniHamed:2009vw}, showing that each actually also possesses the non-manifest superconformal symmetry.

It has been shown very recently in \cite{Drummond:2010qh} how the roles in the Yangian of the original and dual superconformal symmetries can be interchanged via T-duality. In this case, the Yangian generators annihilate the amplitude with the MHV part factored out, rather than the whole amplitude. Therefore there are two equivalent ways to look at the symmetries of scattering amplitudes. The role of twistors and momentum twistors introduced was very important, indeed the representation of the T-dual version of the Yangian in terms of the  momentum twistors is identical to that of the original version in terms of the usual twistors. This T-duality property of the symmetry algebra is the Yangian version of the T-self-duality property of the full AdS${}_5\times$S${}^5$ background of the string sigma-model \cite{Berkovits:2008ic,Beisert:2008iq,Beisert:2009cs}.

The T-dual relationship between twistor and momentum twistor space is exactly the property which appears in the Grassmannian formulas of \cite{ArkaniHamed:2009dn} and \cite{Mason:2009qx}. This property suggests that the Grassmannian formula should be thought of as the most general way of constructing a Yangian invariant. Certainly, the invariance of its form under the T-duality swap between twistors and momentum twistors is a property that such a formula should have. In \cite{Drummond:2010qh} we were also able to show the Yangian invariance of the Grassmannian integral formula directly by simply applying the level-one generators. What we found was that the integrand (i.e. the form being integrated over the Grassmannian) transforms into a total derivative under the Yangian variation. This implies that for any closed contour the result of integration will be a Yangian invariant.

In this paper we show that the form of the Grassmannian integral is uniquely fixed by requiring Yangian invariance. Specifically, we will use the methods developed in \cite{Drummond:2010qh} and demonstrate that one cannot modify the integrand by a non-constant multiplicative function without breaking Yangian invariance. In fact such a modification only has to be constant almost everywhere as it could, in principle, have discontinuities across special hyperplanes in the Grassmannian, specifically where one of the consecutive minors (special Pl\"ucker coordinates) in the Grassmannian integral vanishes.

This paper is structured as follows. In the next section we recall the symmetries of scattering amplitudes, focusing in particular on the Yangian symmetry and its T-dual version. Then in section \ref{invariants} we show how it is possible to construct invariants under $sl(m|m)$. In section \ref{grassmannian} we review  the basic structure of the Grassmannian integrals and their Yangian invariance. In section \ref{uniqueness}, which contains the main result of this paper,  we show the uniqueness of the invariant form. The last section is dedicated to the study of the Yangian invariance when relaxing the homogeneity conditions relevant for $\NN=4$ super Yang-Mills amplitudes.

%This dual superconformal symmetry is broken by loop corrections but becomes exact in the recently proposed Higgs regularisation \cite{Alday:2009zm,Henn:2010bk}.

\section{Scattering amplitudes and Yangian symmetry}

Scattering amplitudes in $\NN = 4$ super Yang-Mills exhibit many remarkable properties. In order to exhibit them simply it is extremely useful to have a manifestly $\mathcal{N}=4$ supersymmetric formulation of the on-shell amplitudes.
Using Grassmann variables $\eta^A$, transforming in the fundamental representation of $su(4)$, we can write the superfield $\Phi$ describing the  on-shell supermultiplet of $\cN=4$ super Yang-Mills theory as
\be
\Phi = G^+ + \eta^A \Gamma_A + \tfrac{1}{2!} \eta^A \eta^B S_{AB} + \tfrac{1}{3!} \eta^a \eta^B \eta^C \e_{ABCD} \overline{\Gamma}^D + \tfrac{1}{4!} \eta^A \eta^B \eta^C \eta^D \e_{ABCD} G^-
\label{onshellmultiplet}
\ee
where $G^+,\Gamma_A,S_{AB}=\tfrac{1}{2}\e_{ABCD}\overline{S}^{CD},\overline{\Gamma}^A,G^-$ are the positive helicity gluon, gluino, scalar, anti-gluino and negative helicity gluon states depending on a light-like momentum $p^{\alpha \dot\alpha}=\lambda^{\alpha} \tilde{\lambda}^{\dot \alpha}$.
The helicity of the superfield is 1 so the amplitude for the scattering of $n$ superfields satisfies the helicity condition for each particle, i.e.
\be
h_i \mathcal{A}(\Phi_1,\ldots,\Phi_n) = \mathcal{A}(\Phi_1,\ldots,\Phi_n), \qquad i=1,\ldots,n
\label{Ahelicity}
\ee
where the helicity operator is
\be
h_i = -\tfrac{1}{2} \lam_i^\a \frac{\del}{\del \lam_i^\a} + \tfrac{1}{2} \tlam_i^\adt \frac{\del}{\del \tlam_i^\adt} + \tfrac{1}{2} \eta_i^A \frac{\del}{\del \eta_i^A}.
\ee
The tree-level amplitudes  can be written as follows,
\be
\mathcal{A}(\Phi_1,\ldots,\Phi_n)=\mathcal{A}_n =  \frac{\d^4(p) \d^8(q)}{\l<12\r> \ldots \l<n1\r>} \cP_n(\lam_i,\tlam_i,\eta_i) = \mathcal{A}_n^{\rm MHV} \mathcal{P}_n
\label{amp}
\ee
where $\mathcal{P}_n$ is a function with no helicity,
\be
h_i \mathcal{P}_n = 0, \qquad i=1,\ldots,n.
\label{Phelicity}
\ee
The superconformality of $\NN = 4$ super Yang-Mills theory is reflected in the structure of scattering amplitudes. In fact the superconformal algebra is generically $su(2,2|4)$ with central charge  $c = \sum_i (1-h_i)$, but imposing the homogeneity condition (\ref{Ahelicity}) the algebra becomes  $psu(2,2|4)$\footnote{We will generically refer to $sl(4|4)$ and indeed $sl(m|m)$ throughout the paper as the issue of the choice of real form does not arise in the considerations we make here.}.
At tree-level, where there are no infrared divergences, amplitudes are therefore annihilated by the generators of the standard superconformal symmetry\footnote{Up to contact terms \cite{Bargheer:2009qu,Korchemsky:2009hm,Sever:2009aa}.}
\be
j_a \mathcal{A}_n = 0, \label{scs}
\ee
where $j_a$ is any generator of  $psu(2,2|4)$,
\be
j_a \in \{p^{\a\adt},q^{\a A}, \bar{q}^{\adt}_A,m_{\a\b}, \bar{m}_{\adt\bdt},r^A{}_B,d,s^\a_A,\bar{s}_{\adt}^A,k_{\a \adt} \}.
\ee
and they are represented by a sum over single particle generators
\be
j_a = \sum_{k=1}^n j_{ka}.
\label{levelzero}
\ee
The invariance was shown directly by applying the generators to the explicit form of the amplitudes in \cite{Witten:2003nn} for MHV amplitudes and \cite{Korchemsky:2009jv} for NMHV amplitudes.

Recently it has been discovered that the amplitudes have also a hidden  symmetry, dual superconformal symmetry \cite{Drummond:2008vq}. This was revealed by introducing a  dual coordinate space related to the on-shell space by the following
\be
x_i^{\a \adt} - x_{i+1}^{\a \adt} = \lam_i^\a \tlam_i^\adt, \qqquad \th_i^{\a A} - \th_{i+1}^{\a A} = \lam_i^\a \eta_i^A.
\label{dualvars}
\ee
Therefore the  amplitudes can be expressed in the dual variables
\be
\mathcal{A}_n = \frac{\d^4(x_1-x_{n+1}) \d^8(\th_1-\th_{n+1})}{\l<12\r>\ldots \l<n1\r>} \cP_n(x_i,\th_i),
\ee
and they turn out to be covariant under certain generators of the dual superconformal algebra.
In particular, if we denote with $J_a$ any generator of the {\sl dual} copy of the superconformal algebra,
\be
J_a \in \{P_{\a\adt},Q_{\a A}, \bar{Q}_{\adt}^A,M_{\a\b}, \overline{M}_{\adt\bdt},R^A{}_B,D,C,S_\a^{A},\overline{S}^{\adt}_A,K^{\a \adt} \},
\ee
the generators giving the covariance are $K$, $S$, $D$ and $C$. By  redefining such generators as in \cite{Drummond:2009fd}, the covariance can be rephrased as an invariance of $\mathcal{A}_n$, so that dual superconformal symmetry becomes simply
\be
J'_a \mathcal{A}_n = 0
\label{dual}
\ee
and
\be
J'_a \in \{P_{\a\adt},Q_{\a A}, \bar{Q}_{\adt}^A,M_{\a\b}, \overline{M}_{\adt\bdt},R^A{}_B,D',C'=0,S_\a^{\prime A},\overline{S}^{\adt}_A,K'^{\a \adt} \}.
\ee

In \cite{Drummond:2009fd} it was shown that the generators $j_a$ together with $S'$ (or $K'$) generate the Yangian of the superconformal algebra, $Y(psu(2,2|4))$. The generators $j_a$ form the level-zero $psu(2,2|4)$ subalgebra\footnote{We use the symbol $[O_1,O_2]$ to denote the bracket of the Lie superalgebra, $[O_2,O_1] = (-1)^{1+|O_1||O_2|}[O_1,O_2]$.},
\be
[j_a,j_b] = f_{ab}{}^{c} j_c.
\ee
The level-one generators $j_a^{(1)}$  are defined by
\be
[j_a,j_b\!{}^{(1)}] = f_{ab}{}^{c} j_c\!{}^{(1)}
\ee
and represented by the bilocal formula,
\be
j_a\!{}^{(1)} = f_{a}{}^{cb} \sum_{k<k'} j_{kb} j_{k'c}.
\label{bilocal}
\ee
The full symmetry of the tree-level amplitudes can be therefore rephrased as
\be
j \mathcal{A}_n = j^{(1)} \mathcal{A}_n = 0.
\ee
Expressed in terms of the twistor space variables $\mathcal{Z}^{\AA} = (\tilde{\mu}^{\alpha}, \tilde\lambda^{\dot\alpha} , \eta^A)$, the level-zero and level-one generators of the Yangian symmetry assume a simple form
\begin{align}
j^{\AA}{}_{\BB} &= \sum_i \cZ_i^{\AA} \frac{\del}{\del \cZ_i^{\BB}}, \label{twistorsconf}\\
j^{(1)}{}^{\AA}{}_{\BB} &= \sum_{i<j} (-1)^{\CC}\Bigl[\cZ_i^{\AA} \frac{\del}{\del \cZ_i^{\CC}} \cZ_j^{\CC} \frac{\del}{\del \cZ_j^{\BB}} - (i,j) \Bigr].
\label{twistoryangian}
\end{align}
Both of the formulas (\ref{twistorsconf}) and (\ref{twistoryangian}) are understood to have the supertrace proportional to $(-1)^{\AA} \delta^{\AA}_{\BB}$ removed. In this representation the generators of superconformal symmetry are first-order operators while the level-one Yangian generators are second order.

In \cite{Drummond:2010qh} it was demonstrated that  there exists an alternative T-dual representation of the symmetry. In this version it is the dual superconformal symmetries $J_a$ which give the level-zero generators, while the standard conformal symmetry together with the dual superconformal symmetry provides the level-one generators. 
In this case, the generators act on the function $\mathcal{P}_n$, where the full MHV tree-level amplitude is factored out.
%As in the previous case, some of the level-one generators have to be modified to annihilate $\mathcal{P}_n$.
The statement of invariance can be written
\be
J_a \mathcal{P}_n = J^{(1)}_a \mathcal{P}_n = 0
\ee
where the level-one generators are given by
\be
J^{(1)}_a = f_{a}{}^{cb} \sum_{i<j} J_{ib} J_{jc}.
\label{dualbilocal}
\ee
It is possible to rewrite the generators in the momentum (super)twistor representation defined in \cite{Hodges:2009hk} $\cW_i^{\AA} = (\lam_i^\a,\mu_i^\adt,\chi_i^A)$. These variables are algebraically related to the on-shell superspace variables $(\lambda,\tilde\lambda,\eta)$ via the introduction of dual coordinates (\ref{dualvars})
and are the twistors associated to this dual coordinate space,
\be
\mu_i^{\dot\alpha} = x_i^{\alpha \dot\alpha} \lambda_{i \alpha}, \qquad \chi_i^A = \theta_i^{\alpha A} \lambda_{i \alpha}.
\ee
These variables linearise {\sl dual} superconformal symmetry.
Written in the momentum twistor representation, the Yangian of dual superconformal symmetry takes an identical form to (\ref{twistoryangian}) up to the change from twistors to momentum twistors,
\begin{align}
J^{\AA}{}_{\BB} &= \sum_i \cW_i^{\AA} \frac{\del}{\del \cW_i^{\BB}},\label{momtwistordsconf}\\
J^{(1)}{}^{\AA}{}_{\BB} &= \sum_{i<j} (-1)^{\CC}\Bigl[\cW_i^{\AA} \frac{\del}{\del \cW_i^{\CC}} \cW_j^{\CC} \frac{\del}{\del \cW_j^{\BB}} - (i,j) \Bigr].
\label{momtwistoryangian}
\end{align}
Again, both formulas are understood to have the supertrace removed.
We recall \cite{Drummond:2009fd} that the formulas (\ref{twistoryangian}) and (\ref{momtwistoryangian}) are actually cyclically symmetric (up to level-zero generators, which are symmetries) even though this property is not obvious from their definitions. This is due to a special property of the superalgebra $sl(4|4)$ which we are studying, namely that it has a vanishing Killing form, a property of all $sl(m|m)$ superalgebras. In fact all the statements which we will make about the construction of invariants under the Yangian $Y(sl(4|4))$ will be equally valid for the case $Y(sl(m|m))$.

\section{Invariants of $sl(m|m)$}
\label{invariants}

We would like to address how it is possible to construct invariants under the Yangian $Y(sl(m|m))$ in the representation (\ref{twistorsconf}), (\ref{twistoryangian}) or equivalently (\ref{momtwistordsconf}), (\ref{momtwistoryangian}). To start with we must address the issue of invariance under the $sl(m|m)$ subalgebra. We will decompose the supertwistor $\cW$ into bosonic and fermionic components,
\be
\cW^{\AA} = (W^{A'},\chi^A).
\ee
It is clear that, since every superconformal generator has a specific Grassmann degree (-1,0 or +1), invariants can be decomposed according to their Grassmann degree (or simply degree). Since $sl(m|m)$ has an $sl(m)$ subgroup that rotates the fermionic variables into each other then this degree must occur in units of $m$. In the $sl(4|4)$ case which interests us most, this is just the expansion of $\NN=4$ super Yang-Mills amplitudes into MHV, NMHV, NNMHV etc.

We will begin with an obvious set of invariants of degree $mk$, familiar from the integrand of the Grassmannian formula of \cite{ArkaniHamed:2009dn},
\be
\prod_{a=1}^k \delta^{m|m}\Bigl(\sum_{i=1}^n t_{ai} \cW_i \Bigr).
\ee
Here the $t_{ai}$ are just some arbitrary complex parameters which define $k$ linear combinations of $n$ supertwistors $\cW_i$. Taking the most general combination of such invariants yields an integral formula
\be
I_k = \int dt f(t) \prod_a \delta^{m|m}\Bigl( \sum_i t_{ai} \cW_i \Bigr).
\label{genform}
\ee
The integration in this formula is actually only $k(n-k)$ dimensional. The reason is that the delta functions do not depend on all $kn$ variables $t_{ai}$, as they have a $GL(k)$ gauge symmetry. When taking linear combinations we should therefore only integrate over the independent degrees of freedom. We will see explicitly that the delta functions naturally arise in a `gauge-fixed' form, dependent only on $k(n-k)$ parameters $t_{ai}$. We lose no generality therefore by taking the function $f(t)$ to be invariant under the $GL(k)$ gauge symmetry.

Now we will see that (\ref{genform}) is the most general form of an $sl(m|m)$ invariant, at least for $k$ small enough. The argument goes as follows. First let us Fourier transform the bosonic variables $W_i^{A'} \rightarrow \tW_{iA'}$, but not the fermionic ones. Half of the fermionic symmetries take the form,
\be
Q_{A'}^{A} = \sum_i \tW_{iA'} \chi_i^A.
\ee
Suppose we look for an invariant of degree zero. We conclude immediately that the only possibility is
\be
I_0(Z,\chi) = \prod_i \delta^m(\tW_i),
\ee
because by assumption we are at degree zero so the invariant cannot vanish for Grassmann reasons. In the original variables $\cW_i$ this is just
\be
I_0(\cW) = 1.
\ee
Now let us look for an invariant of degree $m$. Invariance under $Q_{A'}^A$ can now occur for Grassmann reasons but in order to do so it must be that the Grassmann structure of $Q_{A'}^{A}$ is in fact given only by a single $Q^A$,
\be
Q_{A'}^{A} = \tW^{\rm ref}_{A'} Q^A.
\label{parallel}
\ee
For this to occur it must be that all $\tW_i$ are parallel, i.e. proportional to one of them, $\tW_I$ say (so that $\tW^{\rm ref} = \tW_I$). This implies the invariant takes the form
\be
I_m(\tW,\chi,t) = \Bigl( \prod_{i\neq I} \delta^m(\tW_i - t_{Ii}\tW_I)\Bigr) \delta^m \Bigl(\sum_{i=1}^n t_{Ii} \chi_i \Bigr) X(\tW_I).
\ee
Here $t_{Ii}$ are some constants of proportionality relating each of the $\tW_i$ to $\tW_I$ (by definition $t_{II}=1$).
We have used the delta functions to write any remaining $\tW$ dependence as a function $X$ of the variable $\tW_I$. In fact $X(\tW_I)$ must be a constant because acting with the other half of the fermionic symmetry generators we find
\be
\sum_i \frac{\partial^2}{\partial \tW_{iA'} \partial \chi_i^A} I_m = \Bigl( \prod_{i\neq I} \delta^m(\tW_i - t_{Ii}\tW_I)\Bigr) \partial_A \delta^m \Bigl(\sum_{i=1}^n t_{Ii} \chi_i \Bigr) \frac{\partial X(\tW_I)}{\partial \tW_{IA'}}.
\ee
which only vanishes if $X$ is a constant.

Taking a general linear combination yields the formula
\be
I_m(\tW,\chi) = \int dt f(t) \prod_{i \neq I} \delta^m(\tW_i - t_{Ii}\tW_I) \delta^m(\sum_i t_{Ii} \chi_i).
\ee
Fourier transforming back to $\cW_i$ yields
\be
I_m(\cW) = \int dt f(t) \delta^{m|m}\Bigl(\sum_i t_{Ii} \cW_i \Bigr),
\ee
which is indeed of the form (\ref{genform}) in a fixed $GL(1)$ gauge where $t_{II}=1$.

We can perform a similar argument for invariants of degree $2m$. Now invariance under $Q_{A'}^A$ implies that there are really only two Grassmann combinations $Q_I$ and $Q_J$,
\be
Q_{A'}^A = \tW_{IA'}Q_I^{A} + \tW_{JA'}Q_J^A.
\ee
This means that all $\tW_i$ have to be a linear combination of two independent $\tW$ variables ($\tW_I$ and $\tW_J$ say) and we deduce
\be
I_{2m}(\tW,\chi,t) = \Bigl( \prod_{i\neq I,J} \delta^m(\tW_i - t_{Ii}\tW_I - t_{Ji}\tW_J)\Bigr) \delta^m \Bigl(\sum_{i=1}^n t_{Ii} \chi_i \Bigr) \delta^m \Bigl(\sum_{i=1}^n t_{Ji} \chi_i\Bigr)  X(\tW_I,\tW_J).
\ee
Here we have defined $t_{II}=t_{JJ} =1$ and $t_{IJ}=t_{JI}=0$. As before we find that for invariance under the other fermionic symmetries we require $X$ to be constant. Again, taking a general linear combination yields
\be
I_{2m}(\tW,\chi) = \int dt f(t) \Bigl( \prod_{i\neq I,J} \delta^m(\tW_i - t_{Ii}\tW_I - t_{Ji}\tW_J)\Bigr) \delta^m \Bigl(\sum_{i=1}^n t_{Ii} \chi_i \Bigr) \delta^m \Bigl(\sum_{i=1}^n t_{Ji} \chi_i\Bigr).
\ee
Fourier transforming yields a formula of the general type (\ref{genform}) in a fixed $GL(2)$ gauge where $t_{II}=t_{JJ}=1$ and $t_{IJ}=t_{JI}=0$.

It is clear that we go on with this procedure until we reach $k=m$. Beyond that stage we can no longer argue that the bosonic variables are constrained because any number of $\tW_{iA'}$ can always be expressed as a linear combination of $m$ of them. Thus this argument only works straightforwardly for $k\leq m$. By Fourier transforming the fermionic variables instead of the bosonic ones (so that the other half of the supersymmetry becomes a multiplication operator) we could make the same argument for invariants of degree $(n-k)m$ for $k \leq m$.
Note that our arguments here made no assumption of the regularity or otherwise of the final invariant. Indeed the formula (\ref{genform}) can produce invariants which are regular or singular (i.e. with bosonic delta functions present even after the integrations over the $t_{ai}$). We have assumed that we can Fourier transform the bosonic variables however. In fact this assumption is natural because we actually want invariants which can be Fourier transformed for considering the Grassmannian formulas in the next sections. It is possible however that there may be some invariants not of this form which have no Fourier transform.
We believe however that it is very likely that formula (\ref{genform}) is actually the most general form of all invariants (both regular and singular) for all values of $k$, irrespective of $m$.

Finally let us make a final comment regarding (\ref{genform}). As we have already described it is clear that the delta functions actually possess a $GL(k)$ symmetry acting on the $t_{ai}$ variables. This means that they only really depend on $k(n-k)$ of the integration variables and are independent of the remaining $k^2$. Indeed in the explicit construction we have just presented for $k \leq m$ we found that the delta functions appear in a particular fixed $GL(k)$ gauge with $k^2$ of the $t$ parameters not being integrated over. The only benefit of the extra parameters is to allow different combination of the variables to be vanishing. For example we could not write (\ref{parallel}) with $\tW^{\rm ref} = \tW_I$ if $\tW_I$ itself is vanishing. Of course we can always choose a different $\cW_{I'}$ in that case and we would end up with (\ref{genform}) but in a different $GL(k)$ gauge where $t_{1I'}=1$ instead of $t_{1I}$. Therefore the function $f(t)$ actually only depends on $k(n-k)$ of the integration parameters and we lose no generality by taking $f(t)$ to be invariant under the $GL(k)$ gauge symmetry. Therefore the natural space associated to the invariants of $sl(m|m)$ symmetry is really the Grassmannian $G(k,n)$ which is $k(n-k)$-dimensional\footnote{We thank Nima Arkani-Hamed for discussions of this point.}. The main question which we will address in this paper is to what extent the function $f(t)$ is fixed by imposing invariance under the full Yangian symmetry.

First we will discuss the Grassmannian integral formula of  \cite{ArkaniHamed:2009dn} and the ways Yangian symmetry can be seen in that particular case. In particular we will recall that invariance can be seen as the property that the form being integrated over the Grassmannian transforms into a total derivative under the Yangian symmetry. Then we will argue that in fact this is the {\sl only} form with that property.

\section{Yangian symmetry and the Grassmannian integral}
\label{grassmannian}

Recently a remarkable formula has been proposed which computes leading singularities of scattering amplitudes in the $\NN=4$ super Yang-Mills theory \cite{ArkaniHamed:2009dn}. The formula takes the form of an integral over the Grassmannian $G(k,n)$, the space of complex $k$-planes in $\mathbb{C}^n$. The integrand is a specific $k(n-k)$-form $K$ to be integrated over cycles $C$ of the corresponding dimension, with the integral being treated as a multi-dimensional contour integral.
The result obtained depends on the choice of contour and is non-vanishing for closed contours because the form has poles located on certain hyperplanes in the Grassmannian,
\be
\mathcal{L} = \int_C K.
\ee

The form $K$ is constructed from a product of superconformally-invariant delta functions\footnote{Following recent discussions of the subject we write $\delta^{4|4}(\ldots)$ to represent the $sl(4|4)$ invariant combination of twistor variables. For the bosonic variables which are complex, it is better to think of these delta functions as poles where the integration contour is chosen such that these poles are always enclosed. This point of view was developed in \cite{ArkaniHamed:2009sx} and allows for interesting deformations of the contour which yield alternative expressions for amplitudes.} of linear combinations of twistor variables. It is due to this factor that the integral depends on the kinematic data of the $n$-point scattering amplitude of the gauge theory. The delta functions are multiplied by a cyclically invariant function on the Grassmannian which has poles. Specifically the formula takes the following form in twistor space
\be
\mathcal{L}_{\rm ACCK}(\cZ) = \int \frac{ D^{k(n-k)}c}{\MM_1 \ldots \MM_n} \prod_{a=1}^k \delta^{4|4}\Bigl(\sum_{i=1}^n c_{ai} \cZ_i\Bigr) ,
\label{ACCK}
\ee
where the $c_{ai}$ are complex parameters which are integrated choosing a specific contour.
The denominator is the cyclic product of  consecutive $(k \times k)$ minors $\MM_p$  made from the columns $
p,\ldots,p+k-1$ of the $(k \times n)$ matrix of the $c_{ai}$
\beq
\MM_{p} \equiv (p~p+1~p+2\ldots p+k-1)  .
\eeq 	
As described in \cite{ArkaniHamed:2009dn} the formula (\ref{ACCK}) has a $GL(k)$ gauge symmetry which implies that $k^2$ of the $c_{ai}$ are gauge degrees of freedom and therefore should not be integrated over. The remaining $k(n-k)$ are the true coordinates on the Grassmannian. Therefore the measure $D^{k(n-k)}c$ is a form of degree $k(n-k)$. It is locally $GL(k)$ invariant and globally $GL(n)$ invariant and is given explicitly in terms of the $c_{ai}$ in \cite{Mason:2009qx}.

This formula (\ref{ACCK}) produces leading singularities of N${}^{k-2}$MHV scattering amplitudes when suitable closed integration contours are chosen. There is also a momentum space version \cite{ArkaniHamed:2009vw},
\be
\label{ACCKmom}
\mathcal{L}_{\rm ACCK} (\lambda,\tilde\lambda,\eta) = \int \frac{{D^{k(n-k)}c} \prod_a d^2 \rho_a}{\MM_1 \ldots \MM_n} \prod_{a=1}^k \delta^{2}\Bigl(\sum_{i=1}^n c_{ai} \tilde \lambda_i\Bigr)\delta^4\Bigl(\sum_{i-1}^n c_{ai} \eta_i \Bigr) \prod_{i=1}^n \delta^2\Bigl(\lambda_i - \sum_{a=1}^k \rho_{a} c_{ai}\Bigr) ,
\ee
where $k$ auxiliary spinors are introduced as integration variables to represent the chiral half of the bosonic delta functions from (\ref{ACCK}). Strictly speaking the two formulas (\ref{ACCK}) and (\ref{ACCKmom}) are only related to each other so simply in (2,2) signature where all the bosonic variables are real. In Minkowski signature the passage between twistor space and momentum space is more involved. In particular it requires a choice of integration contours and is not as straightforward as the simple Fourier transform between (\ref{ACCK}) and (\ref{ACCKmom}).

The formula (\ref{ACCK}) has T-dual version \cite{Mason:2009qx}, expressed in terms of momentum twistors. The momentum twistor Grassmannian formula takes the same form as the original
\be
\mathcal{L}_{\rm MS} = \int \frac{D^{k(n-k)}t}{\MM_1 \ldots \MM_n} \prod_{a=1}^k  \delta^{4|4}\Bigl(\sum_{i=1}^n t_{ai} \cW_i\Bigr) ,
\label{MS}
\ee
but now it is the dual superconformal symmetry that is manifest. The integration variables $t_{ai}$ are again a $(k \times n)$ matrix of complex parameters. The formula (\ref{MS}) produces the same objects as (\ref{ACCKmom}) but now with the MHV tree-level amplitude factored out. They therefore contribute to N${}^k$MHV amplitudes.

The equivalence of the two formulations (\ref{ACCKmom}) and (\ref{MS}) was shown in \cite{ArkaniHamed:2009vw}, through a change of variables. Therefore, since each of the formulas has a different superconformal symmetry manifest, they both possess an invariance under the Yangian $Y(psl(4|4))$. The Yangian symmetry of these formulas was explicitly demonstrated in \cite{Drummond:2010qh} by directly applying the Yangian level-one generators to the Grassmannian integral itself. This yields a total derivative
\be
J^{(1)}{}^{\AA}{}_{\BB} K = d \Omega^{\AA}{}_{\BB}
\label{exactvar}
\ee
which guarantees that $\mathcal{L}$ is invariant when the contour is closed. To fix notation we work with momentum twistors and the form of the Yangian given in equations (\ref{momtwistordsconf}) and (\ref{momtwistoryangian}). This allows us to work with a manifestly $sl(4|4)$ invariant notation without having to worry about performing the integral transform from (\ref{ACCK}) back into momentum space. In particular this means that the analysis is valid directly for Minkowski signature spacetime.
Of course if we are just interested in twistor space then our analysis concerning invariance is equally valid for (\ref{ACCK}) using the twistor space formulas (\ref{twistorsconf}) and (\ref{twistoryangian}). We could also have worked with the formula (\ref{ACCKmom}) but at the cost of not having a manifestly $sl(4|4)$ covariant presentation of the symmetry generators.

%%%%%%%%%%%%%%%%
%%%%%%%%%%%%%%%%
\section{Uniqueness of the invariant form}
\label{uniqueness}

Here we would like to see if it is possible to modify the form in some way that preserves the property (\ref{exactvar}), i.e. that its level-one variation is a total derivative. In order to do this we imagine writing the same integral as before but now with an extra arbitrary function $f(t)$ on the Grassmannian in the integrand,
\be
\tilde{\mathcal{L}}_{n,k} = \int \frac{D^{k(n-k)}t}{\MM_1 \ldots \MM_n} f(t) \prod_a \delta_a.
\label{modL}
\ee
In (\ref{modL}) we have used the shorthand notation for the delta functions\footnote{To be general we now write $\delta^{m|m}$ for the delta functions invariant under $sl(m|m)$.},
\be
\delta_a = \delta^{m|m}\Bigl(\sum_{i=1}^n t_{ai} \cW_i \Bigr).
\ee

After performing the integrations, the $t_{ai}$ become functions of the bosonic twistor variables. We are only interested here in functions of the $\cW_i$ and not their conjugates $\overline{\cW}_i$. The bosonic delta functions (really they are poles inside integration contours) impose relations between the $W_i$ (the bosonic parts of the $\cW_i$) and the $t_{ai}$. Of course they also relate the $\overline{W}_i$ to $\bar{t}_{ai}$ so we will take the function $f(t)$ to be a function only of the $t_{ai}$ and not of $\bar{t}_{ai}$.

We will also impose the constraints on $f(t)$ that $\tilde{\mathcal{L}}$ is homogeneous of degree zero in each of the $\cW_i$ separately,
\be
\cW_i^{\AA} \frac{\partial}{\partial \cW_i^{\AA}} \tilde{\mathcal{L}} = 0, \qquad i=1,\ldots,n.
\label{homconditions}
\ee
These conditions are none other than the helicity conditions (\ref{Ahelicity}) in the momentum twistor language.
They imposes $n$ first-order conditions on the function $f(t)$. Indeed acting with the $\cW$-scaling operator on the delta functions we see that they can be exchanged for $t$-scaling operators in the following way,
\be
\cW_i^\CC \frac{\partial}{\partial \cW_i^\CC} \longrightarrow \OO_i = \sum_{a=1}^k t_{ai} \frac{\partial}{\partial t_{ai}}.
\label{OOi}
\ee
The constraints on the function $f(t)$ which guarantee homogeneity are then simply
\be
\OO_i f(t) = 0.
% \quad i=k+1,\ldots ,n, \qqquad \UU_i f(t) = 0, \quad i=1,\ldots ,k.
\label{homtvars}
\ee
These constraints simply reflect the fact that, for the integral to be homogeneous in the supertwistors $\cW_i$, the $t_{ai}$ each have to scale oppositely to $\cW_i$ so that the delta functions are homogeneous. The integrand then has to compensate the weights of the measure under these induced scalings. The denominator factor $\MM_1 \ldots \MM_n$ exactly compensates the transformation of the measure. Therefore the factor $f(t)$ must itself be invariant under the scalings generated by $\OO_i$. Later we will see what happens when we relax the requirement of homogeneity of degree zero in all $\cW_i$.

In \cite{Drummond:2010qh} when we applied the level-one generator to $\mathcal{L}$ we found that there is an induced first-order transformation of the $t$ variables, generated by $k$ independent operators $\OO_b^{\AA}$.
Acting with the level-one Yangian generators on $\tilde{\mathcal{L}}$ we find a similar expression to that in \cite{Drummond:2010qh}, except that the factor $f(t)$ is present in the integrand,
\be
\tfrac{1}{2} J^{(1)}{}^{\AA}{}_{\BB} \tilde{\mathcal{L}}_{n,k} = \sum_b \int \frac{D^{k(n-k)}t\, f(t)}{\MM_1 \ldots \MM_n} \bigl[ \OO_b^{\AA} - \mathcal{V}_b^{\AA} \bigr] \partial_{\BB} \delta_b \prod_{a\neq b} \delta_a.
\label{variation}
\ee
Here $\mathcal{V}_b^{\AA}$ is simply a linear combination of supertwistors given by the following triangular double sum,
\be
\mathcal{V}_b^{\AA} = \sum_{i<j} t_{bi} \cW_i^{\AA}.
\label{VbA}
\ee
The other term in (\ref{variation}) contains the operator $\OO_b^{\AA}$ that induces a transformation of the $t_{ai}$. This operator is defined as follows,
\be
\OO_b^\AA = \sum_{i<j} \cW_i^{\AA} \OO_{ij} t_{bj},
\label{ObA}
\ee
where $\OO_{ij}$ is built from the generators of $gl(n)$ transformations of the $t_{ai}$,
\be
\OO_{ij} = \sum_{a = 1}^k t_{ai} \frac{\partial}{\partial t_{aj}} .
\label{OOij}
\ee
This form for the operator follows simply from acting with $J^{(1)}{}^{\AA}{}_{\BB}$ on the delta functions in (\ref{modL}). We refer the reader to \cite{Drummond:2010qh} for the details of the derivation.
This precise form of the operator already assumes that we have imposed the homogeneity constraints (\ref{homconditions}) or (\ref{homtvars}). When we consider the case where we do not impose the homogeneity conditions we will see that the operator is slightly different.

In \cite{Drummond:2010qh}, in order to show the property (\ref{exactvar}) in the case when $f$ is a constant function, we commuted the operator $\OO_b^{\AA}$ back past the minors in the denominator and found that
\be
\Bigl[\frac{1}{\MM_1 \ldots \MM_n} , \OO_b^{\AA} \Bigr] = \frac{\mathcal{V}_b^{\AA}}{\MM_1 \ldots \MM_n}.
\ee
In other words,
the transformation of the denominator $\MM_1 \ldots \MM_n$ under the operators $\OO_b^{\AA}$ was exactly what was required to compensate the multiplicative contribution $\mathcal{V}_b^{\AA}$ from the level-one variation (\ref{variation}).

When we insert a non-trivial function $f(t)$ into the integrand as in (\ref{modL}) we must require that it is invariant under the action of $\OO_b^{\AA}$, in order to preserve the property that the variation of the integrand is a total derivative,
\be
[f(t),\OO_b^{\AA}] = 0.
\ee
We require invariance under the action $\OO_b^{\AA}$ for each value of $b$ separately because each term in the sum over $b$ is independent due to the derivative acting on one of the delta functions. We will see this explicitly later when we fix the gauge.

The precise form of the operators $\OO_b^{\AA}$ will be important for our discussion, but for now we just need to note that they are linear in the $\cW_i$. The delta functions impose $k$ conditions among the $\cW_i$ which may be used to eliminate, say, the first $k$ of them\footnote{One of the delta functions in the integrand comes with a derivative $\partial_\BB$. The constraint imposed by the delta function can still be used to eliminate the corresponding $\cW_b^{\AA}$ as the only thing that arises in commuting $\cW^\AA$ past the derivative is proportional to the supertrace $(-1)^\AA \delta^{\AA}_{\BB}$ which can then be dropped as we recall that the operator $J^{(1)}{}^{\AA}{}_{\BB}$ is actually understood to be supertraceless.} so there are at best only $k(n-k)$ independent components $O_{bi}$,
\be
\OO_b^{\AA}  = \sum_{i=k+1}^n \cW_i^{\AA} O_{bi}.
\ee
These independent components define $k(n-k)$ vector fields $O_{bi}$ on the Grassmannian.
This implies $k(n-k)$ first-order conditions on the form of the function $f(t)$, one solution of which is that $f(t)$ is a constant. The question of whether this is sufficient to fix $f(t)$ to be constant is the question of whether the vector fields are generically linearly independent or not. If they become linearly dependent only on certain planes on the Grassmannian then we will deduce that $f(t)$ has to be a constant almost everywhere.
To determine whether the $k(n-k)$ vector fields are linearly independent we can simply evaluate the measure form on them. We can think of this quantity as a determinant. We will define 
\be
\det O = (\MM_1)^k {D^{k(n-k)}t}(O_{1 k+1},...,O_{kn}).
\ee
The factor of $(\MM_1)^k$ is there to ensure that the degrees of homogeneity are all equal.
If $\det O$ vanishes we know that there are linear dependencies among the vector fields. Of course since $\det O$ is a function on the Grassmannian it will in general have zeros located at certain places. The important point for understanding whether the symmetries fix the form $K$ is whether or not $\det O = 0$ {\sl everywhere} on the Grassmannian.
The quantity we have defined $\det O$ is clearly independent of any particular coordinate choice on the Grassmannian. It is also cyclic since the operator $J^{(1)}{}^{\AA}{}_{\BB}$ is cyclic and the factor of $(\MM_1)^k$ exactly compensates for a cyclic rotation with respect to the choice of expanding in terms of the last $(n-k)$ of the $\cW_i$. Note that we have assumed nothing about the function $f(t)$, this is simply a statement about the operator $J^{(1)}{}^{\AA}{}_{\BB}$.
As we will now describe, in order to actually compute the quantity $\det O$ it is very convenient to fix a gauge and perform the calculation in that gauge.

Now we will perform the explicit calculation to show that the determinant we are interested in is indeed generically invertible. We will work in the gauge where the first $k$ columns are fixed to be the identity matrix so that the delta functions take the form,
\be
\delta_a = \delta^{4|4}\Bigl( \cW_a + \sum_{l=k+1}^n t_{al} \cW_l \Bigr)
\ee
and the measure is simply the wedge product of all the $dt_{ai}$,
\be
{D^{k(n-k)}t}= \prod_{a=1}^k \prod_{i=k+1}^n dt_{ai}.
\ee
In this gauge the $k(n-k)$ coordinates on the Grassmannian are simply the remaining $t_{ai}$ for $i=k+1, \ldots n$.

As we have described, in the proof of invariance of the unmodified Grassmannian formula in \cite{Drummond:2010qh} we found that the level-one variation induced a particular transformation of the integration parameters $t$. In the gauge we are working in, the operator $\OO_b^{\AA}$ must be written in such a way as to take into account the specific gauge-fixing.
In the fixed gauge, acting with the level-one Yangian generators on $\tilde{\mathcal{L}}$ we find a similar expression to that in \cite{Drummond:2010qh}, except that again the factor $f(t)$ is present in the integrand,
\be
\tfrac{1}{2} J^{(1)}{}^{\AA}{}_{\BB} \tilde{\mathcal{L}}_{n,k} = \sum_b \int \frac{d^{k(n-k)}t f(t)}{\MM_1 \ldots \MM_n} \bigl[ \NN_b^{\AA} - \mathcal{V}_b^{\AA} \bigr] \partial_{\BB} \delta_b \prod_{a\neq b} \delta_a.
\label{Nvariation}
\ee
Here $\mathcal{V}_b^{\AA}$ is as before in (\ref{VbA}). The other term in (\ref{variation}) contains the gauge-fixed version of $\OO_b^{\AA}$ which we write as $\NN_b^{\AA}$,
\be
\NN_b^\AA = \sum_{i<j} \cW_i^{\AA} \NN_{ij} t_{bi}.
\label{NbA}
\ee
Here $\NN_{ij}$ is built from the operators,
\be
\OO_{ij} = \sum_{a = 1}^k t_{ai} \frac{\partial}{\partial t_{aj}} \text{ and } \UU_{ij}  = \sum_{l=k+1}^n t_{jl}\frac{\partial}{\partial t_{il}}.
\label{OOijUUij}
\ee
Specifically we have
\be
\NN_{ij} = - \UU_{ij} \text { if } j \leq k \text{ and } \NN_{ij} = \OO_{ij} \text{ if } j>k.
\label{NNij}
\ee
We refer the reader to \cite{Drummond:2010qh} for the derivation of the form of $\NN_{b}^{\AA}$.

As we have argued, to preserve the property that the integrand transforms into a total derivative we need,
\be
[\, \NN_b^{\AA} , f(t) ] = 0
\label{fcomm}
\ee
for each value of $b$ separately because these variations are independent. We can verify this explicitly in this gauge as follows. First we note that due to invariance under the level-zero symmetry we only need to show invariance under one of the level-one generators. This is sufficient as then commutators with the level-zero generators will generate the rest of the Yangian. As we only need to show invariance for one of the level-one generators we can take the index $\BB$ to correspond to a fermionic variable and the index $\AA$ to correspond to a bosonic one. Then expanding the Grassmann delta functions one finds only one term in the sum over $b$ which contains, for example, $(\chi_1)^4 \ldots (\chi_{k-1})^4$ (this would be the term where $b=k$). The other terms all have one of the relevant $\chi$ removed due to the fermionic derivative $\partial_{\BB}$.
Therefore this term must vanish on its own. The same can be argued for all the values of $b$.

The conditions (\ref{fcomm}) are simply the following first order partial differential equations,
\be
\sum_{i<j} \cW_i^{\AA} t_{bi} \NN_{ij} f(t) = 0,
\label{pdes}
\ee
with $\NN_{ij}$ as described in (\ref{OOijUUij}) and (\ref{NNij}). We need to determine whether there is any solution for $f$ other than a constant. As we described we will now proceed to extract the $k(n-k)$ independent equations from (\ref{pdes}) by looking at the coefficients of the independent $\cW_i$ which we will take to be $\cW_{k+1}, \ldots , \cW_n$.

Splitting the sum over $i$ in (\ref{pdes}) into the parts $i\leq k$ and $i>k$ we find that the operator acting on $f$ can be written,
\begin{align}
&\sum_{i=1}^k \sum_{j=i+1}^n \cW_i^{\AA} t_{bj} \NN_{ij} + \sum_{i=k+1}^{n-1} \sum_{j=i+1}^n \cW_i^{\AA} t_{bj} \NN_{ij}  \notag \\
&= - \sum_{l=k+1}^n \sum_{i=1}^k \sum_{j=i+1}^n  t_{il} \cW_l^\AA t_{bj} \NN_{ij} + \sum_{l=k+1}^{n-1} \sum_{j=l+1}^n \cW_l^{\AA} t_{bj} \NN_{lj}  ,
\label{Noperator}
\end{align}
where in the first term we have eliminated all $\cW_i$ for $1 \leq i \leq k$ using the delta functions in the integral (\ref{variation}) and in the second we have renamed the summation variable $i \rightarrow l$.

Since the remaining $\cW_l$ for $k+1 \leq l \leq n$ are independent we can look at the coefficient of $\cW_l$ in (\ref{Noperator}) which we will call $N_{bl}$,
\be
N_{bl} = -\sum_{i=1}^k \sum_{j=i+1}^n t_{il} t_{bj} \NN_{ij} + \sum_{j=l+1}^n t_{bj} \NN_{lj}.
\ee
Writing this in terms of $\UU$ and $\OO$ from (\ref{NNij}) we find
\begin{align}
N_{bl} =  \sum_{i=1}^{k-1} \sum_{j=i+1}^k t_{il} t_{bj} \UU_{ij} - \sum_{i=1}^k \sum_{j=k+1}^n t_{il} t_{bj} \OO_{ij} + \sum_{j=l+1}^n t_{bj} \OO_{lj}.
\end{align}
Explicitly in terms of derivatives with respect to the $t$ variables this becomes
\be
N_{bl} = \sum_{i=1}^{k-1} \sum_{j=i+1}^k t_{il} t_{bj} \sum_{m=k+1}^n t_{jm} \frac{\partial}{\partial t_{im}} - \sum_{i=1}^k \sum_{j=k+1}^n t_{il} t_{bj} \frac{\partial}{\partial t_{ij}} + \sum_{j=l+1}^n t_{bj} \sum_{a=1}^k t_{al} \frac{\partial}{\partial t_{aj}}.
\ee
So we find that
\be
N_{bl} = \sum_{a,j} O_{bl,aj} \frac{\partial}{\partial t_{aj}}
\ee
where the $k(n-k) \times k(n-k)$ matrix $O$ is given by
\be
O_{bl,aj} = \sum_{r=a+1}^k t_{al} t_{br} t_{rj} - t_{al} t_{bj} + \delta(j>l) t_{bj}t_{al}.
\ee
The first term simplifies because it vanishes if $a\geq b$ and contributes to only one term if $a<b$. So we have
\begin{align}
O_{bl,aj} &=  \delta(a<b) t_{al} t_{bj} - t_{al} t_{bj} + \delta(j>l) t_{bj} t_{al} \notag \\
&= -\delta(a \geq b) t_{al} t_{bj} + \delta(j>l) t_{al} t_{bj} \notag \\
&= [\delta(j>l) - \delta(a \geq b)] t_{al} t_{bj}.
\label{canonMform}
\end{align}
%In fact the matrix $M$ is ambiguous because we also have the conditions (\ref{homtvars}). This means that, for example, we can add to $M_{bl,aj}$ terms of the form $\delta_{ab} t_{al} t_{bj}$ or $\delta_{jl} t_{al} t_{bj}$.

In the gauge we are working in, the quantity $\det O$ we need to investigate is simply the determinant of the matrix $O$.
First let us note that it is a polynomial of degree $2k(n-k)$ in the $t_{ai}$ variables. Then we note that if we regard the matrix $O$ as an $(n-k)\times(n-k)$ array of $k\times k$ blocks then we see that $t_{1,k+1}$ appears in the every entry in the first row and every entry in the first column and in fact the entry at the very top left of the matrix is $t_{1,k+1}^2$. This means that the determinant has a factor of $t_{1,k+1}^2$. In the gauge we are working in $t_{1,k+1}$ is just the value of the minor $\MM_2$. Since this determinant could be calculated gauge-invariantly this means that in general the determinant has a double zero at $\MM_2=0$. The operator $J^{(1)}{}^{\AA}{}_{\BB}$ we are acting with is cyclic so this means that $\det O$  must have a double zero where any of the cyclically related minors $\MM_p$ vanish. Therefore we conclude that $\det O$ has a factor of $[\MM_1 \ldots \MM_n]^2$. In our fixed gauge this is already a polynomial of the correct degree and so we conclude
\be
\det O = c(k,n-k) [\MM_1 \ldots \MM_n]^2,
\ee
for some overall factor $c(k,n-k)$ which is independent of the $t_{ai}$. We write the arguments of $c$ as $k$ and $(n-k)$ because the roles of $k$ and $(n-k)$ can clearly be interchanged without altering the general form of the matrix (one can equally well regard it as $k \times k$ array of $(n-k) \times (n-k)$ blocks). Therefore with this definition $c$ is symmetric in its arguments.

To show that the matrix is generically invertible, all we need to do is show that $c(k,n-k) \neq 0$ for all $n$ and $k$. We say generically invertible because of course it becomes non-invertible on special hyperplanes in the Grassmannian where one of the minors $\MM_p$ vanishes.

In fact it is simple to see that $c(k,n-k)=1$ for all $k,n$. We note that it is clearly true for $k=1$ since the matrix is triangular in that case and the determinant is obviously $[t_2 \ldots t_n]^2 = [\MM_1 \ldots \MM_n]^2$. Then we can assume for an induction that $c(k,n-k)=1$ for all $k$ up to and including $k=K$. Now we consider the case $k=K+1$. If $(n-k)<k$ we have $c(k,n-k) = c(n-k,k)$ and this is equal to one by the inductive assumption. If $(n-k)>k$ then we can set the first $k$ columns of the matrix of the $t_{ai}$ to be the $k \times k$ identity matrix. Then the matrix $O$ simplifies so that the only pieces contributing to the determinant are a $k^2 \times k^2$ block of unit determinant at the top left and a $k(n-2k)\times k(n-2k)$ block at the bottom right which is of the same form as $O$. Thus we find $c(k,n-k)=c(k,n-2k)$. We can repeat the process of subtracting $k$ from the second argument of $c$ until we can swap the arguments in order to reduce the value of the first one and then we conclude that $c(k,n-k)=1$ by the induction on $k$.

To summarise we have seen that the conditions (\ref{pdes}) can be written as follows,
\be
\sum_{a,j} O_{bl,aj} \frac{\partial}{\partial t_{aj}} f(t) = 0,
\label{Mdf=0}
\ee
where the matrix $O$ satisfies
\be
\det O = [\MM_1 \ldots \MM_n]^2.
\label{main}
\ee
Since the matrix $O$ is generically invertible we can multiply (\ref{Mdf=0}) by the inverse of $O$ and deduce that $f(t)$ must be constant almost everywhere. In principle the function $f(t)$ can have discontinuities across the hyperplanes defined by the vanishing of the minors $\MM_p$ but the only continuous function allowed is a constant function.

\section{Invariants of non-zero homogeneities}

We will now discuss what happens when we do not insist on imposing the homogeneity requirements (\ref{homconditions}) relevant for superamplitudes in $\NN =4$ super Yang-Mills theory\footnote{We thank Nima Arkani-Hamed and Freddy Cachazo for providing an explicit example for $k=1,n=6$ of a new invariant with non-zero homogeneities which inspired us to investigate this possibility.}. It is necessary to return to the form of the level-one Yangian generators,
\be
J^{(1)}{}^{\AA}{}_{\BB} = \sum_{i<j} \biggl[(-1)^\CC \cW_i^\AA \frac{\partial}{\partial \cW_i^\CC} \cW_j^\CC \frac{\partial}{\partial \cW_j^{\BB}} -(i,j) \biggr]
= \biggl( \sum_{i<j} - \sum_{j<i} \biggr)\biggl[(-1)^\CC \cW_i^\AA \frac{\partial}{\partial \cW_i^\CC} \cW_j^\CC \frac{\partial}{\partial \cW_j^{\BB}} \biggr].
\ee
We recall that the operator should be understood to have the supertrace on the indices $\AA$ and $\BB$ removed. The level-one generators can be simplified by writing the sum where $j<i$ as
\be
\sum_{j<i} = \sum_{i,j} - \sum_{i=j} - \sum_{i<j}.
\ee
The term where $i$ and $j$ are summed independently becomes proportional to the level-zero generators acting on $\mathcal{L}$ and so it vanishes. The sum where $i=j$ becomes
\be
\sum_i (-1)^\CC \cW_i^\AA \frac{\partial}{\partial \cW_i^\CC} \cW_i^\CC \frac{\partial}{\partial \cW_i^\BB} = \sum_i \biggl[ \cW_i^\AA \frac{\partial}{\partial \cW_i^\BB} \cW_i^{\CC} \frac{\partial}{\partial \cW_i^\CC} - \cW_i^\AA \frac{\partial}{\partial \cW_i^\BB} \biggr].
\ee
The second term on the RHS can be dropped as it is a level-zero generator. The first term can also be dropped if, as is the case when $f$ is a constant, all the degrees of homogeneity are equal. If they are not equal however, the first term on the RHS contributes to the variation.
As we will see, this has the consequence that if one does not impose the homogeneity conditions then it is possible to find non-constant functions $f$ in (\ref{modL}) which are consistent with symmetry under the Yangian $Y(sl(4|4))$.

As we have seen, the $\cW$-scaling operators can be exchanged for $t$-scaling operators when acting on the delta functions. In the gauge where we fix the first $k$ columns of the matrix of $t_{ai}$ to be the $k \times k$ identity matrix, the replacement is as follows,
\be
\cW_i^\CC \frac{\partial}{\partial \cW_i^\CC} \longrightarrow \NN_i = (\OO_i , - \UU_i).
\ee
Here we have defined
\be
\OO_i = \sum_{a=1}^k t_{ai} \frac{\partial}{\partial t_{ai}} \text{ for }i>k  \qquad \UU_i = \sum_{l=k+1}^n t_{il} \frac{\partial}{\partial t_{il}} \text{ for } i\leq k.
\label{OOiUUi}
\ee
This has the effect that the level-one variation of the integral is modified with respect to (\ref{Nvariation}),
\be
\tfrac{1}{2} J^{(1)}{}^{\AA}{}_{\BB} \tilde{\mathcal{L}}_{n,k} = \sum_b \int \frac{d^{k(n-k)}t f(t)}{\MM_1 \ldots \MM_n} \bigl[ \tilde{\NN}_b^{\AA} - \mathcal{V}_b^{\AA} \bigr] \partial_{\BB} \delta_b \prod_{a\neq b} \delta_a.
\label{Ntvariation}
\ee
The operator $\tilde{\NN}_b^{\AA}$ contains an extra term compared with (\ref{NbA}),
\be
\tilde{\NN}_b^{\AA} = \NN_b^{\AA} + \hat{\NN}_b^{\AA},
\label{tildeNbA}
\ee
where
\be
\hat{\NN}_b^{\AA} = \sum_i \cW_i^{\AA} \NN_i t_{bi}.
\ee
For invariance we require the conditions
\be
[\, \tilde{\NN}_b^{\AA} , f(t) ] = 0.
\label{tildeinvariance}
\ee
Proceeding the same way as in the previous section we find that the extra term in the variation translates into a modification of the matrix $O$. Specifically we find that the conditions (\ref{tildeinvariance}) are expressed as the following first-order partial differential equations,
\be
\sum_{a,j} \tilde{O}_{bl,aj} \frac{\partial}{\partial t_{aj}} f(t) = 0.
\ee
The matrix $\tilde{O}$ contains an extra piece compared with $O$ from equation (\ref{canonMform}),
\begin{align}
\tilde{O}_{bl,aj} &= [\delta(j>l) - \delta(a\geq b) + \tfrac{1}{2}(\delta_{ab} + \delta_{jl})] t_{al}t_{bj} \notag\\
&= \tfrac{1}{2}[\delta(j>l) - \delta(l>j) - \delta(a>b) + \delta(b>a)]t_{al} t_{bj}.
\end{align}
We see that $\tilde{O}$ is an antisymmetric matrix.
As before we can see that the determinant of $\tilde{O}$ has a double zero at $t_{1,k+1}=0$. Thus we deduce, as before that
\be
\det \tilde{O} = \tilde{c}(k,n-k) [\MM_1 \ldots \MM_n]^2.
\ee
Now we have a different overall factor $\tilde{c}(k,n-k)$ which is not always non-zero. Indeed we can see from the antisymmetry of $\tilde{O}$ that, for example, when $k(n-k)$ is odd we have $\tilde{c}(k,n-k)=0$ and hence $\tilde{O}$ is not invertible anywhere on the Grassmannian in this case. Indeed it is possible to find eigenvectors with vanishing eigenvalue which are given by the gradient of some function $f(t)$. An example is in the case $k=1$, $n=6$, where we find
\be
\sum_{a,j} \tilde{O}_{bl,aj} \frac{\partial}{\partial t_{aj}} \Bigl(\frac{t_{11} t_{13} t_{15}}{t_{12} t_{14} t_{16}} \Bigr) = 0.
\ee
As can be seen very nicely using the arguments of \cite{ArkaniHamed:2009vw}, T-duality relates the case $k=3$, $n=6$ to the $k=1$, $n=6$ one. For $k=3$, $n=6$ the corresponding invariant function is
\be
f(t) = \biggl( \frac{(123)(345)(561)}{(234)(456)(612)}\biggr).
\ee
The invariants constructed here can not contribute to scattering amplitudes in $\NN=4$ super Yang-Mills theory as they would have different helicity assignments. Very optimistically one might imagine that there is some other theory with a $Y(sl(m|m))$ symmetry whose amplitudes are described by these invariants. However it might be that, while such invariants exist, it is impossible to build a consistent set of amplitudes from them.

%%%%%%%%%%%%%%%%

\section{Conclusions}

In this paper we have analysed the uniqueness of the Grassmannian formulas of \cite{ArkaniHamed:2009dn,Mason:2009qx}. 
Following the approach of \cite{Drummond:2010qh}, we have demonstrated that the only Yangian invariant form of the relevant degree on the Grassmannian is the one of \cite{ArkaniHamed:2009dn}. The form cannot be deformed by a non-constant function $f$ without spoiling Yangian invariance. Indeed our main result eq. (\ref{main}), shows that the only continuous function allowed is a constant function though in principle $f$ can have discontinuities across special hyperplanes in the Grassmannian where one of the $\MM_p$ minors vanishes. This shows the uniqueness of the Grassmannian integrals when demanding Yangian invariance and the zero homogeneity conditions.
We have also seen that, relaxing the requirement of zero degrees of homogeneity, it is possible to find deformations of the Grassmannian formula maintaining the Yangian invariance.

%%%%%%%%%%%%%%%%

\subsection*{Note Added}
In \cite{KSpaper} a different approach to the problem of the uniqueness of the Grassmannian integral is presented.

\subsection*{Acknowledgements}

We would like to thank Nima Arkani-Hamed, Marco Bill\`{o}, Freddy Cachazo, Laurent Gallot and Emery Sokatchev for many useful discussions.
This research was supported in part by the French Agence Nationale de la Recherche under grant ANR-06-BLAN-0142 and CNRS.
JMD would like to thank the Institute of Advanced Study in Princeton, where part of this work was carried out, for warm hospitality. LF would like to thank the Humboldt University in Berlin for hospitality during the final stages of this work.

%%%%%%%%%%%%%%%%%%%%%%%%%%%%%%%%%%%%%%%%%%%%%%%%%

\section*{Some generalities on $gl(m|m)$ and its Yangian}

We will begin with the defining representation of $gl(m|n)$.
We define $E^{\AA}{}_{\BB}$ to be an
$(m|n) \times (m|n)$
matrix with a 1 in the entry in row $\AA$ and column $\BB$ and 0 everywhere else. The matrix satisfies the product
\be
E^{\AA}{}_{\BB} E^{\CC}{}_{\DD} = \delta^\CC_\BB E^{\AA}{}_{\DD},
\ee
from which follows the commutation relations of $gl(m|n)$,
\be
[E^{\AA}{}_{\BB},E^{\CC}{}_{\DD}] = \delta^\CC_\BB E^{\AA}{}_{\DD} - (-1)^{(\AA+\BB)(\CC+\DD)} \delta^\AA_\DD E^{\CC}{}_{\BB} = f^{\AA}{}_{\BB}{}^{\CC}{}_{\DD}{}_{\EE}{}^{\FF} E^{\EE}{}_{\FF},
\ee
where the structure constants $f$ are given by
\be
f^{\AA}{}_{\BB}{}^{\CC}{}_{\DD}{}_{\EE}{}^{\FF} E^{\EE}{}_{\FF} = \delta_\BB^\CC \delta_\EE^\AA \delta_\DD^\FF - (-1)^{(\AA+\BB)(\CC+\DD)} \delta_\DD^\AA \delta_\EE^\CC \delta_\BB^\FF.
\ee
If we remove the supertrace from the generators $E^{\AA}{}_{\BB}$ then we have the algebra $sl(m|n)$. In the case where $m=n$ we can also remove the trace, leading to $psl(n|n)$.

One can define a metric on $gl(m|n)$ by taking the supertrace of the product of two generators in the fundamental representation,
\be
g^{\AA}{}_{\BB}{}^{\CC}{}_{\DD} = {\rm str}[E^{\AA}{}_{\BB} E^{\CC}{}_{\DD}] = (-1)^\AA \delta^\CC_\BB \delta^\AA_\DD.
\ee
The inverse metric is then
\be
(g^{-1}){}_{\AA}{}^{\BB}{}_{\CC}{}^{\DD} = (-1)^\BB \delta^\DD_\AA \delta^\BB_\CC.
\ee
We can define `raised' structure constants as
\be
f^{\AA}{}_{\BB}{}_{\GG}{}^{\HH}{}_{\EE}{}^{\FF} = f^{\AA}{}_{\BB}{}^{\CC}{}_{\DD}{}_{\EE}{}^{\FF} (g^{-1})_{\CC}{}^{\DD}{}_{\GG}{}^{\HH} = (-1)^\GG(\delta^\HH_\BB\delta^\AA_\EE\delta^\FF_\GG - (-1)^{(\AA+\BB)(\AA+\EE)}\delta^\AA_\GG \delta^\HH_\EE \delta^\FF_\BB).
\ee
The representation of most interest to us the twistor (or oscillator) representation,
\be
J^{\AA}{}_{\BB} = \cW^\AA \frac{\del}{\del \cW^\BB}.
\ee
It is simple to see that this satisfies the right commutation relations,
\be
[J^{\AA}{}_{\BB},J^{\CC}{}_{\DD}] = \delta_\BB^\CC J^{\AA}{}_{\DD} - (-1)^{(\AA+\BB)(\CC+\DD)} \delta^\AA_\DD J^{\CC}{}_{\BB}.
\ee
For multi-particle invariants we take the sum over single particle representations,
\be
J^\AA{}_{\BB} = \sum_i J_i^\AA{}_{\BB}=\sum_i \cW^\AA_i \frac{\del}{\del \cW_i^\BB}.
\ee
The Yangian generators are given by the bilocal sum,
\be
J^{(1)}{}^{\AA}{}_{\BB} = \sum_{i<j}(-1)^\CC [J_i^{\AA}{}_{\CC} J_j^{\CC}{}_{\BB} - J_j^{\AA}{}_{\CC} J_i^{\CC}{}_{\BB}].
\label{Yangiangenerators}
\ee
They are consistent with cyclicity (i.e. invariant up to terms which are proportional to a generator of the original superalgebra) for those algebras with vanishing Killing form \cite{Drummond:2009fd}. The simple Lie superalgebras which satisfy this condition were classified by Kac \cite{Kac:1977em} and include $psl(m|m)$. It also holds for the central extension $sl(m|m)$ but not for $gl(m|m)$. This can be seen by considering the difference of the definition (\ref{Yangiangenerators}) with that which one obtains by cyclically rotating by one step. Explicitly, the only term which is not proportional to a level-zero algebra generator is the supertrace of (\ref{Yangiangenerators}).

\end{document}